\documentclass{ws-mpla}
\usepackage{graphicx}
\usepackage{comment}
\usepackage[super,sort]{natbib}

\begin{document}
\title{THE STAR FORMATION HISTORIES OF EARLY-TYPE GALAXIES:
INSIGHTS FROM THE REST-FRAME ULTRAVIOLET}
\author{\footnotesize SUGATA KAVIRAJ}
\address{Denys Wilkinson Building, Keble Road, University of
Oxford, UK\\skaviraj@astro.ox.ac.uk}
\maketitle


\begin{abstract}
Our current understanding of the star formation histories of
early-type galaxies is reviewed, in the context of recent
observational studies of their ultra-violet (UV) properties.
Combination of UV and optical spectro-photometric data indicates
that the bulk of the stellar mass in the early-type population
forms at high redshift ($z>2$), typically over short timescales
($<1$ Gyr). Nevertheless, early-types of all luminosities form
stars over the lifetime of the Universe, with most luminous
($-23<M(V)<-21$) systems forming 10-15\% of their stellar mass
after $z=1$ (with a scatter to higher value), while their less
luminous ($M(V)>-21$) counterparts form 30-60\% of their mass in
the same redshift range. The large scatter in the (rest-frame) UV
colours in the redshift range $0<z<0.7$ indicates widespread
low-level star formation in the early-type population over the
last 8 billion years. The mass fraction of young ($<1$ Gyr old)
stars in luminous early-type galaxies varies between 1\% and 6\%
at $z\sim0$ and is in the range 5-13\% at $z\sim0.7$. The
intensity of recent star formation and the bulk of the UV colour
distribution is consistent with what might be expected from minor
mergers (mass ratios $\lesssim$ 1:6) in a $\Lambda$CDM cosmology.
\end{abstract}


\section{Introduction}
The star formation histories (SFHs) of early-type galaxies have
been the subject of intense and controversial debate in modern
astrophysics. The classical `monolithic collapse' hypothesis for
early-type evolution followed the model of Eggen et al. for the
formation of the Galaxy\citet{ELS62}. Refined and implemented by
others \citep{Larson74,Chiosi2002}, this model postulated that
stellar populations in early-type galaxies form in short, highly
efficient starbursts at high redshift ($z\gg1$) and evolve purely
passively thereafter. The \emph{optical} properties of the
early-type population and, in particular, their strict obedience
to simple scaling relations are remarkably consistent with such a
simple, largely empirical hypothesis. The small scatter in the
early-type `Fundamental Plane' \citep{Jorg1996,Saglia1997} and its
apparent lack of evolution with look-back time
\citep{Forbes1998,Peebles2002,Franx1993,Franx1995,VD1996}, coupled
with the homogeneity and lack of redshift evolution in their
optical colours
\citep{BLE92,Bender1997,Ellis97,Stanford98,Gladders98,VD2000} is
strong evidence that the bulk of the stellar population in
early-type galaxies forms at high redshift. Furthermore, the
over-abundance of $\alpha$ elements in these systems
\citep{Thomas1999} indicates that the star formation timescales
are shorter than the typical timescales for the onset of Type Ia
supernovae (SN). For example, if Type Ia progenitors largely
explode within $\sim1$ Gyr of a starburst \citep{Greggio1983},
then the bulk of the star formation in these galaxies probably
takes place on timescales shorter than a Gyr.

While it is consistent with the optical properties of early-type
galaxies, a monolithic SFH does not sit comfortably within the
currently accepted $\Lambda$CDM galaxy formation paradigm in which
the stellar mass in local early-type systems is thought to
accumulate over the lifetime of the Universe. Following the
seminal work of Toomre\citet{Toomre_mergers}, who showed that most
galaxy collisions end in rapid merging and postulated the
formation of spheroidal systems as end-products of such merger
activity, the mechanics of galaxy interactions
\citep{Barnes1992a,Barnes1992b,Hernquist1993} and their link to
the formation of early-type systems \citep{Barnes1996,Bender1996}
have been studied in considerable detail. Modern `semi-analytical'
models of galaxy formation, within the $\Lambda$CDM paradigm,
create early-types through `major mergers', where the mass ratio
of the merging progenitors is 3:1 or lower. The constituent
stellar mass of early-type galaxies is predicted to form both
quiescently in their progenitors and in efficient starbursts when
these progenitors
merge\citep{K96,K98,Cole2000,Hatton2003,Khochfar2003}.

While theoretical arguments may be compelling, the strongest
evidence for the role of interactions in shaping early-type galaxy
evolution and inducing coincident star formation comes from
observation, both in the local Universe and at high redshift. 40\%
of local ellipticals contain dust lanes \citep{Sadler1985}, while
$\sim75$\% contain nuclear dust and by implication gas
\citep{Tomita2000,Tran2001}. The gas is often kinematically
decoupled from the stars, indicating, at least in part, an
external origin, e.g. through the accretion of a gas rich
satellite \citep{sauron5}. Up to two-thirds of nearby early-types
contain shells, ripples and morphological disturbances
\citep{Malin83,VD2005} and a significant fraction exhibit
kinematically decoupled cores \citep{sauron2}, both of which are
evidence for interactions in the recent past. While the detection
of such spatially resolved fine structure is possible only for
galaxies in our local neighbourhood, clear signatures of recent
star formation have been found in individual early-type systems
out to modest redshifts
\citep{Trager2000a,Trager2000b,Fukugita2004}.

Deep ground and space-based imaging are increasingly providing
access to galaxy populations over the last ten billion years of
look-back time. A significant fraction ($\sim$30\%) of luminous
early-type systems at high redshift ($0.4<z<0.8$) exhibit blue
cores, indicative of merger-driven starbursts triggered by the
accretion of low-mass gas-rich companions
\citep{Ferreras2005,Menanteau2001a}. Furthermore, most blue cores
are predominantly contained in early-type galaxies
\citep{Ferreras2005} and, not unexpectedly, they are typically
accompanied by spectral lines characteristic of recent star
formation \citep{Ellis2001}. A fundamental consequence of
early-type evolution in the $\Lambda$CDM model is the gradual loss
of late-type progenitors \citep{VD2001a,Kaviraj2005a} and a
corresponding increase in the fraction of early-type galaxies.
Numerous observational studies have detected such an evolution in
the morphological mix of the Universe. While $\sim$80 percent of
galaxies in the cores of local clusters have early-type morphology
\citep{Dressler80}, a higher fraction of spiral (blue) galaxies
have been reported in clusters at high redshift
\citep{BO84,Dressler97,Couch98,VD2000,Margoniner2001,Andreon2004,Borch2006,Bundy2006},
combined with increased rates of merger and interaction events
\citep{VD99}. Similar results have been found in large-scale
survey data which suggest that the mean mass density on the red
sequence (which is dominated by early-type systems at all
redshifts) has at least doubled since $z=1$ (e.g. Bell et al. 2004
\citep{Bell2004}; Faber et al. 2007 \citep{Faber2007}). A
significant body of observational evidence thus indicates that the
SFH of \emph{at least} some early-type galaxies, and perhaps the
early-type population as a whole, deviates significantly from the
expectations of the classical monolithic collapse hypothesis, both
in terms of their structural evolution and the star formation
experienced by them over the lifetime of the Universe.

Although the majority of early-type studies in the past have
focussed on the optical spectrum, a significant drawback of
optical photometry is its lack of sensitivity to moderate amounts
of \emph{recent star formation} (RSF). While red optical colours
do imply a high-redshift formation epoch for the bulk of the
stellar mass in early-type systems, the optical spectrum remains
largely unaffected by the \emph{minority} of stellar mass that is
expected to form in these objects at low and intermediate
redshifts\citep{Kaviraj2005a}. As a result, it is difficult to
resolve early-type SFHs over the last 8 billion years (where the
predictions of the two competing models diverge the most) using
optical colours alone.

An ideal route to quantifying early-type SFHs at low and
intermediate redshift is to exploit a sensitive indicator of RSF.
If the early-type population could be studied, over a large range
redshift, using such an indicator, then the steady accumulation of
stellar mass could be robustly quantified and strong constraints
applied to the predictions of galaxy formation models over the
last 8 billion years of look-back time. Spectroscopic indicators
of RSF already exist, such as the commonly used $H_{\beta}$ index,
higher order Balmer lines such as $H_{\gamma}$ and $H_{\delta}$
and the D4000 break. With the advent of large spectroscopic
surveys at low redshift, such as the Sloan Digital Sky Survey
(SDSS)\citep{SDSSDR4}, it is possible to employ spectroscopic line
indices to study the local galaxy population. However, equivalent
data at intermediate redshift - certainly on the same scale -
remains scant, although this is gradually changing with efforts
such as the DEEP2 survey\citep{Davis2003}.

\begin{figure}
\begin{center}
\includegraphics[width=0.8\textwidth]{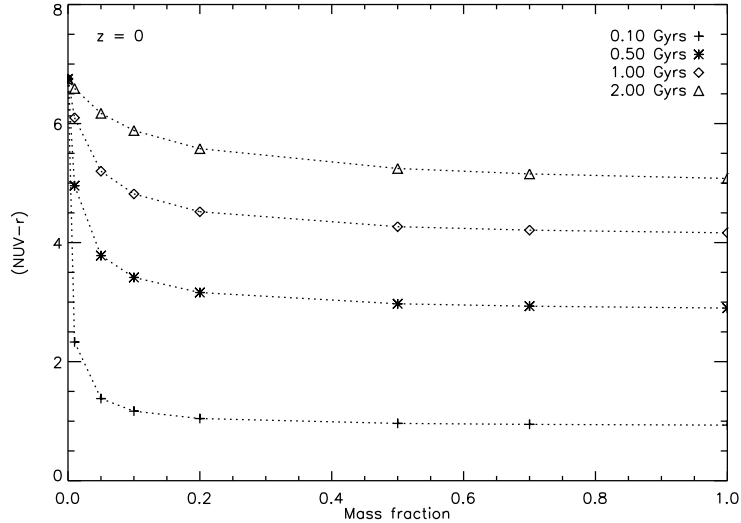}
\caption{The sensitivity of the UV to young stars. We assume two
instantaneous bursts of star formation, where the first burst is
fixed at $z=3$ and the second burst is allowed to vary in age and
mass fraction. The near-UV (NUV) colour of the composite stellar
population is plotted as a function of the age (symbol type) and
mass fraction (x-axis) of the second burst. It is apparent that
even a small mass fraction ($\sim$1\%) of young stars ($\sim0.1$
Gyrs old) causes a dramatic change in the $NUV-r$ colour compared
to what might be expected from a purely old stellar population
($NUV-r\sim 6.8$).}
\label{fig:uv_sense}
\end{center}
\end{figure}

Rest-frame UV photometry provides an attractive alternative. While
its impact on the optical spectrum is relatively weak (and
virtually undetectable, given typical observational and
theoretical uncertainties), a small mass fraction ($<3$\%) of
young ($<1$ Gyr old) stars strongly affects the rest-frame UV
shortward of $3000\AA$. Furthermore, the UV remains largely
unaffected by the age-metallicity degeneracy \citep{Worthey1994}
that typically plagues optical analyses \citep{Kaviraj2006c},
making it an ideal \emph{photometric} indicator of RSF. The
sensitivity of the UV to young stars is demonstrated in Figure
\ref{fig:uv_sense}. We assume two instantaneous bursts of star
formation, where the first burst is fixed at $z=3$ and the second
burst is allowed to vary in age and mass fraction. The near-UV
(NUV; $2300\AA$) colour of the composite stellar population is
plotted as a function of the age (symbol type) and mass fraction
(x-axis) of the second burst. It is apparent that even a very
small mass fraction ($\sim$1\%) of young stars ($\sim0.1$ Gyrs
old) causes a dramatic change in the $NUV-r$ colour compared to
what might be expected from a purely old stellar population
($NUV-r\sim 6.8$). Given that typical observational uncertainties
in the $NUV-r$ colours from modern instrumentation are $\sim0.2$
mag, the usefulness of the UV in detecting residual amounts of RSF
becomes quite apparent.

The advent of the GALEX UV space telescope \citep{Martin2005} and
deep optical surveys (which can be used to trace the rest-frame UV
spectrum at high redshift) has provided us, over the last few
years, with an unprecedented opportunity to quantify the SFHs of
early-type galaxies over the last 8 billion years by exploiting
their (rest-frame) UV properties. This review describes the
results of recent efforts that incorporate UV photometry to study
the evolution of the early-type galaxy population across the
redshift range $0<z<1$. While the emphasis is on studies that have
quantified the buildup of stellar mass in these systems - without
necessarily exploring the processes that drive this star formation
- possible sources of RSF are explored, both in the context of
monolithic collapse and in the standard $\Lambda$CDM cosmology.


\section{The UV colours of nearby early-type galaxies}
\subsection{Early work and evidence for recent star formation}
Ferreras \& Silk \cite{Ferreras2000} were one of the first to
study the rest-frame UV colours of early-type galaxies in the
nearby Universe. Using Hubble Space Telescope (HST) $F300W$ and
optical imaging, they explored the UV colour-magnitude relation
(CMR) of the early-type population in the Abell 851 cluster at
$z\sim0.4$. They found that the slope and, in particular, the
large scatter in the UV colours was consistent with some
early-types having $\sim10$\% of their stellar mass in stars
younger than $\sim500$ Myrs. Detailed modelling of this data
\citep{Ferreras2002} indicated that minor bursts of RSF -
superimposed on an underlying population that forms at high
redshift - leads to a natural explanation of the large scatter
observed in the UV CMR.

Analysis of the rest-frame UV properties of early-type galaxies at
lower redshifts ($z<0.2$) is complicated by the fact that their UV
spectrum may contain contributions from \emph{both} young and old
($>9$ Gyrs old) stellar populations. Core helium burning stars on
the evolved horizontal branch (HB), thought to be the primary
cause of the `UV upturn' phenomenon in massive elliptical galaxies
\citep{Yi97,Yi99}, emit efficiently in the UV. Thus, the
potential contamination of the UV spectrum from such evolved
stellar populations has to be taken into account, before the
contribution from young stellar populations can be gauged.
Clearly, this complicates the robust detection of young stars in
local early-type galaxies.

Large-scale UV data from the GALEX space telescope, unprecedented
in quality and quantity, provides a unique opportunity to study
the RSF-sensitive UV emission from a statistically large sample of
nearby early-type galaxies across a range of luminosities and
environments. Yi et al.\cite{Yi2005} and Rich et
al.\cite{Rich2005} were the first to study the UV emission of the
general early-type population in the local Universe, using GALEX
detections of early-type galaxies drawn from the SDSS
\citep{Bernardi2003d}. To account for contamination from potential
UV upturn, Yi et al. compared the UV spectral slope of each
early-type object in their sample to the spectral energy
distribution (SED) of one of the strongest nearby UV-upturn
galaxies (NGC 4552). They found that no more than 4 out of 62
early-types were consistent with significant amounts of UV upturn
flux and that \emph{at least} 15\% of the early-type population
showed strong signatures of low-level RSF, where 1\%-2\% of the
stellar mass fraction was probably comprised of stars less than a
Gyr old.


\subsection{The GALEX-SDSS view of the local early-type
population} The preliminary result of Yi et al. was refined and
the UV properties of early-type galaxies studied more thoroughly
in the context of galaxy formation models by Kaviraj et
al.\citep{Kaviraj2006b} (K06 hereafter). A sample of early-type
galaxies was drawn from the SDSS by first extracting galaxies that
have largely `de Vaucouleurs' profiles (by setting
\texttt{fracdev}$>0.95$ when selecting objects from the SDSS
database), followed by visual inspection to remove late-type
contaminants. Since scattered light from Active Galactic Nuclei
(AGN) may contaminate the UV spectrum (and thus affect the
estimation of parameters from the SED), AGN were removed using a
`BPT' type analysis\citep{Baldwin1981,Kauffmann2003} using optical
emission line ratios computed from their SDSS spectra and radio
luminosities measured by the FIRST survey. The sample was
restricted to $r<16.8$ and $z<0.11$ to ensure the robustness of
the morphological classification and the detection completeness of
the red sequence, and cross-matched with GALEX data in the Medium
Imaging Survey (MIS) mode with a depth of $m_{AB}\sim23$. GALEX
offers two photometric filters - far-UV (FUV), with an effective
wavelength of $\sim1500\AA$ and the near-UV (NUV), with an
effective wavelength of $\sim2300\AA$. Since UV-upturn flux peaks
in the FUV filter, the focus of this study was the NUV filter.

\begin{figure}
\begin{center}
\includegraphics[width=0.8\textwidth]{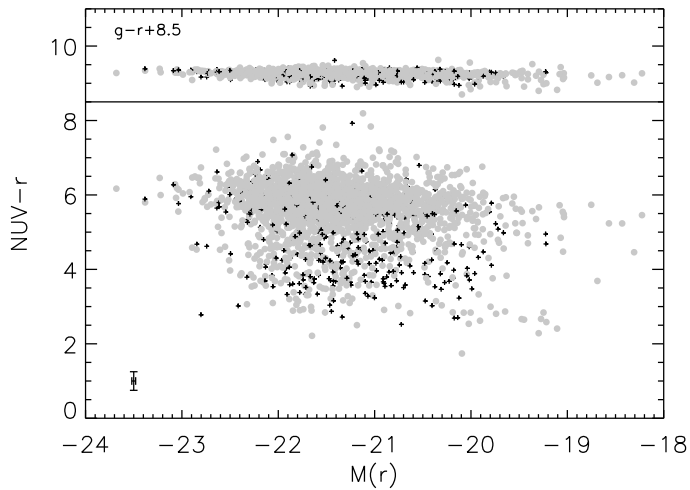}
\caption{The optical $g-r$ (top) and the $NUV-r$ (bottom)
colour-magnitude relations of the local ($0<z<0.11$) early-type
population (filled grey circles). The optical CMR is shown on the
same scale as its UV counterpart, to highlight the significant
difference in their respective scatters. Small black crosses
represent galaxies with active (Type 2) AGN, identified using
optical or radio analyses and consequently removed from the
analysis.} \label{fig:opt_nuv_cmr}
\end{center}
\end{figure}

The basic result of this work is shown in Figure
\ref{fig:opt_nuv_cmr}. The small scatter ($\sim 0.05$ mag) of the
optical CMR (top panel) is in stark contrast to the broadness of
its UV counterpart, which shows a spread of almost 5 mags in
$NUV-r$. An immediate conclusion from Figure \ref{fig:opt_nuv_cmr}
is that the UV colours seem inconsistent with the population as
whole forming exclusively at high redshift. Indeed, if all
early-type galaxies were dustless, simple stellar populations
forming at high redshift (as has frequently been assumed from
their optical colours), the NUV CMR could be expected to have a
spread of less than 1 mag (assuming a spread in metallicities) -
several factors less than what is observed. In light of the
expected behaviour of the $NUV-r$ colour (shown in Figure
\ref{fig:uv_sense}), it is natural to suspect that at least some
of the scatter in the NUV CMR shown in Figure
\ref{fig:opt_nuv_cmr} must be due to the presence of young stars,
especially in the \emph{bluest} early-type galaxies.

To quantify the presence of young stars, K06 parametrised the SFHs
of each early-type object in their sample using a simple model,
where two instantaneous bursts of star formation were assumed to
describe the evolution of each early-type galaxy. Since the
optical colours of the early-type population consistently suggest
that the bulk of the star formation should have taken place at
high redshift, the initial burst was fixed at $z=3$, with the
second burst allowed to vary in age ($t_{YC}$) and mass fraction
($f_{YC}$). A realistic spread in metallicity and dust content
were assumed and marginalised values of $t_{YC}$ and $f_{YC}$ and
their associated uncertainties were extracted. A negligible
fraction of early-types were found to be consistent with an
absence of star formation within the last 2 Gyrs. However, 30\% of
the early-type population, which were also the bluest in the NUV
CMR ($NUV-r<5.5$), showed \emph{unambiguous} signs of RSF
i.e. the 95\% confidence contours in ($t_{YC}$,$f_{YC}$) space
were well-constrained to young ages and mass fractions of a few
per cent. The remaining ($\sim70$\%) galaxies in the sample were
consistent with both small levels of RSF and old ($>2$ Gyr old)
stellar populations, which could not be distinguished, given the
uncertainties in the photometry and the stellar models employed in
the parametrisation.

\begin{figure}
\begin{center}
$\begin{array}{cc}
\includegraphics[width=0.5\textwidth]{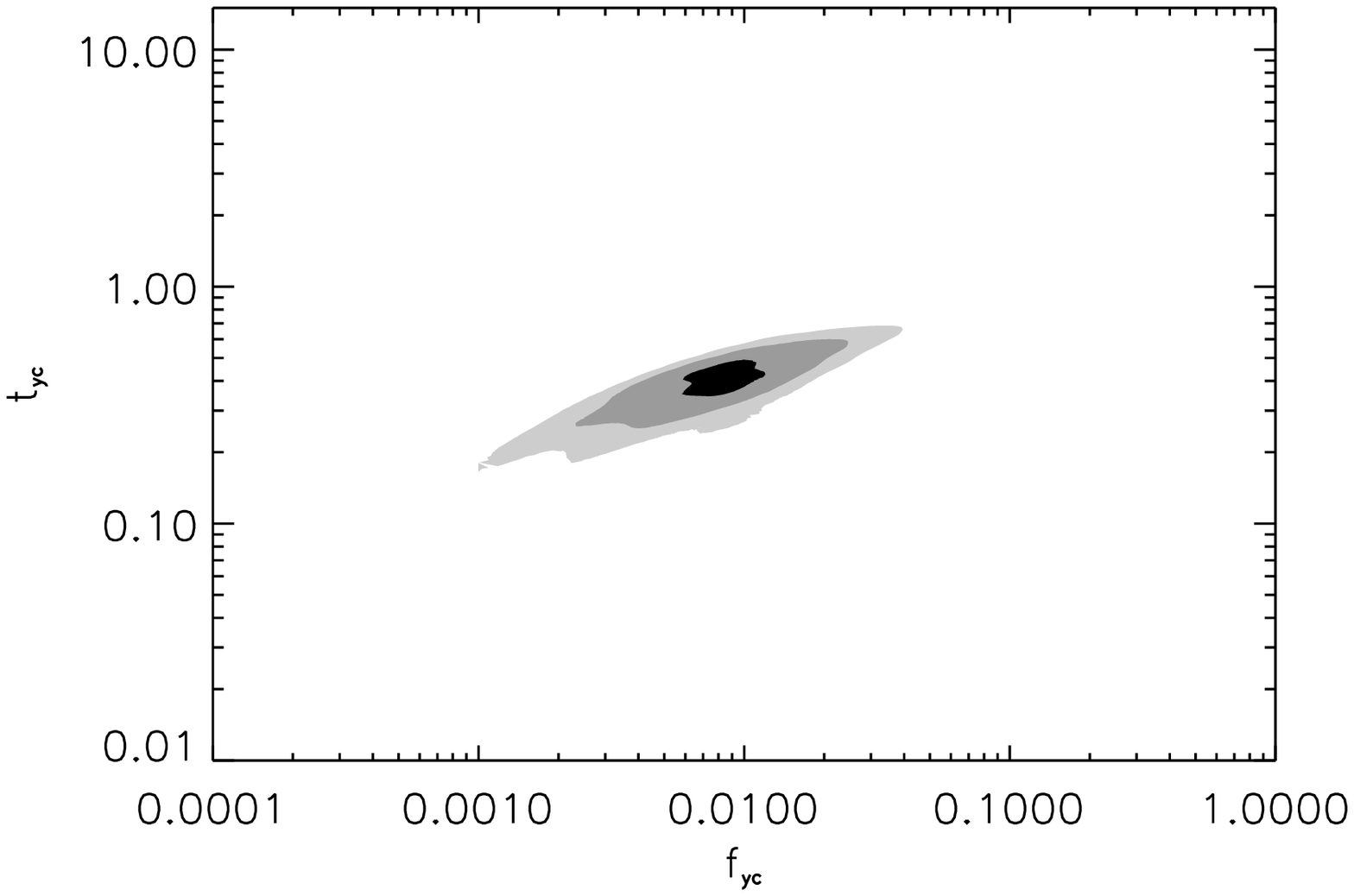}
\includegraphics[width=0.5\textwidth]{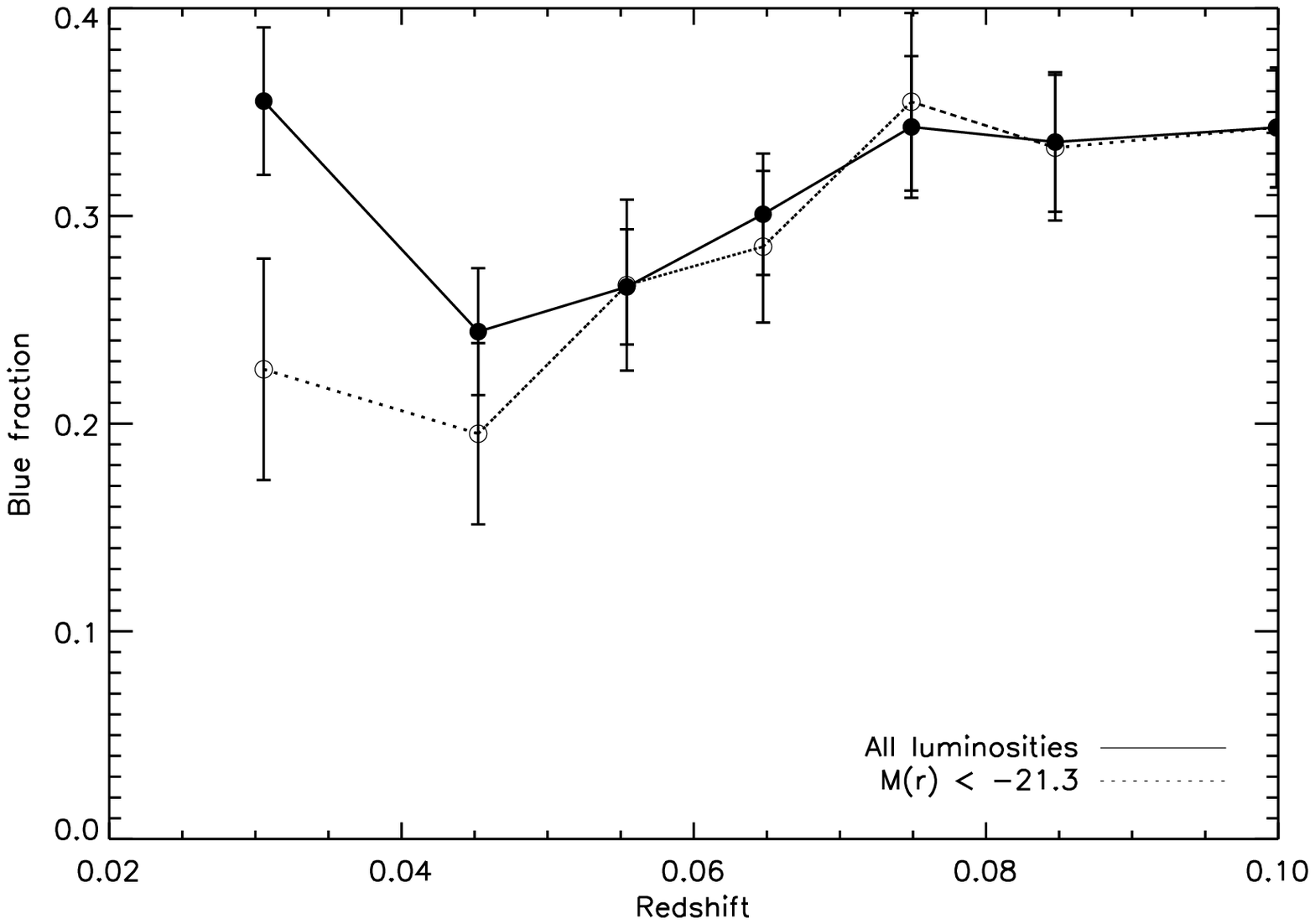}
\end{array}$
\caption{Stacked likelihood map (left panel) of the blue
early-type galaxies in the ($t_{YC}$,$f_{YC}$) parameter space and
the fraction of such blue (star-forming) early-type galaxies
across the redshift range sampled by K06.}
\label{fig:blue_galaxies}
\end{center}
\end{figure}

Figure \ref{fig:blue_galaxies} shows a stacked likelihood map
(left panel) of the blue early-type galaxies in the
($t_{YC}$,$f_{YC}$) parameter space. The typical stellar mass
fraction contributed by young stars in these galaxies is around
1\%, with a typical (mass-weighted) age of $\sim0.5$ Gyrs. The
fraction of such blue early-type galaxies (right panel) is around
30\% and does not fluctuate appreciably (within the errors) in the
redshift range sampled by K06 ($0<z<0.11$).


\subsection{A comparison to models}
While a simple SFH with two instantaneous bursts can be used to
measure the deviation from a monolithic model, it does not reflect
the more complex stellar assembly expected in the $\Lambda$CDM
paradigm. K06 demonstrated (see their Figure 19) that the `blind'
UV predictions of a semi-analytical model, that is calibrated to
the optical colours of the (cluster) early-type population,
provides quantitative agreement with the observed GALEX-SDSS UV
colours of the local early-type population, given reasonable
assumptions for the dust properties and lifetimes of birth clouds
which host the youngest stars. This suggests that the mix of SFHs
that can be predicted in the $\Lambda$CDM framework are reasonably
consistent with those present in the observed early-type
population.

Given the low level of RSF in local early-type galaxies, it is
natural to ask whether the level of cold gas required to supply a
trickle of young stars could plausibly be supplied in a
traditional monolithic scheme. While monolithic models form the
bulk of the stellar mass at high redshift (e.g. $z<2$), the stars
created in this primordial burst would recycle a fraction of their
mass back into the ISM through stellar winds and supernova ejecta.
A fraction or all of this internally recycled gas could
potentially fuel further star formation. Using standard chemical
enrichment prescriptions, K06 studied \emph{hypothetical}
monolithic scenarios, calibrated to reproduce the red optical
colours and high alpha-enhancements of massive early-type galaxies
in the local Universe. They concluded that monolithic scenarios
with reasonable assumptions for the star formation and chemical
enrichment history are unable to produce blue early-type galaxies,
although they might plausibly reproduce some of the galaxies on
the UV red sequence. Note that these scenarios are simply
empirical constructs and do not simulate the dynamical evolution
of accreted gas in a dark matter (DM) halo. While numerical
simulations of baryonic infall in \emph{non-rotating} DM halos can
produce galaxies that fit the optical data of luminous early-type
systems \citep{Chiosi2002}, it is not clear whether the short
accretion timescales required can be achieved in a more realistic
scenario where the infalling material will have some angular
momentum.


\section{UV colours of the high-redshift early-type population:
evidence for stellar mass assembly over the last 8 billion years}
The confirmation of widespread RSF in the local early-type
population and the identification of a sizeable population whose
SFHs cannot be explained through monolithic collapse is an
important first step towards understanding their evolution in the
local Universe. However, a comprehensive understanding of
early-type evolution over a large range in look-back time requires
these results to be extended to high redshift by studying the
rest-frame UV properties of distant galaxy populations.

Kaviraj et al.\citep{Kaviraj2007a} (K07 hereafter) selected a
suite of three recent optical surveys (MUSYC \citep{Gawiser2006},
COMBO-17 \citep{Wolf2004} and GEMS \citep{Rix2004}) in the
well-studied `Extended Chandra Deep Field South' (ECDF-S) that
provide the ideal tools for studying the rest-frame UV photometry
of the high redshift ($0.5<z<1$) galaxy population. The MUSYC
survey offers deep UBVRIzJK imaging of ECDF-S, to AB depths of
$U,B,V,R = 26.5$ and $K = 22.5$. COMBO-17 provides accurate
photometric redshifts through 17 filter photometry, while GEMS
provides $V/z$-band HST (ACS) imaging of galaxies in ECDF-S out to
$z\sim1$, from which morphologies can be deduced through visual
inspection.

\begin{figure}
$\begin{array}{cc}
\includegraphics[width=0.5\textwidth]{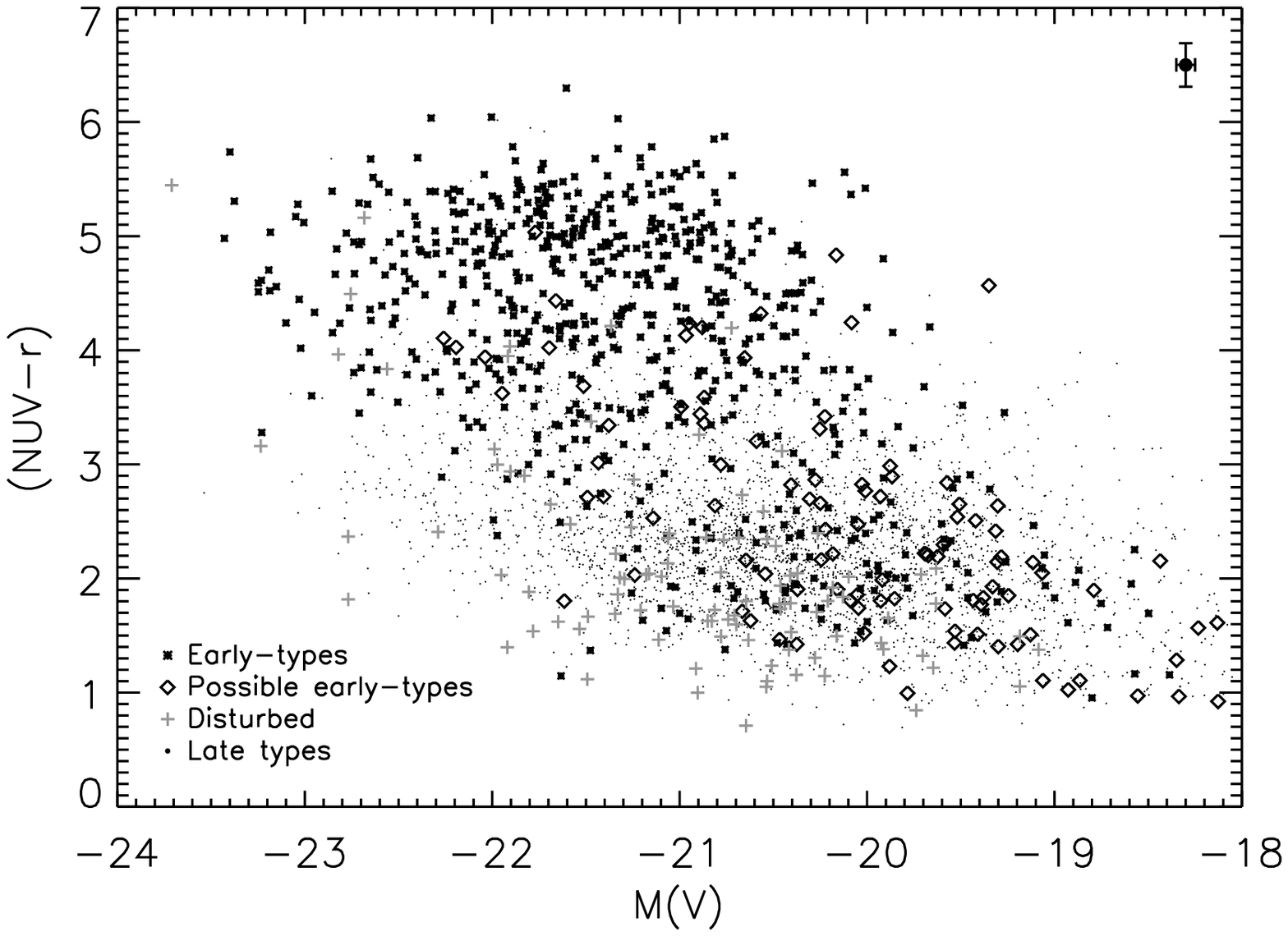} &
\includegraphics[width=0.5\textwidth]{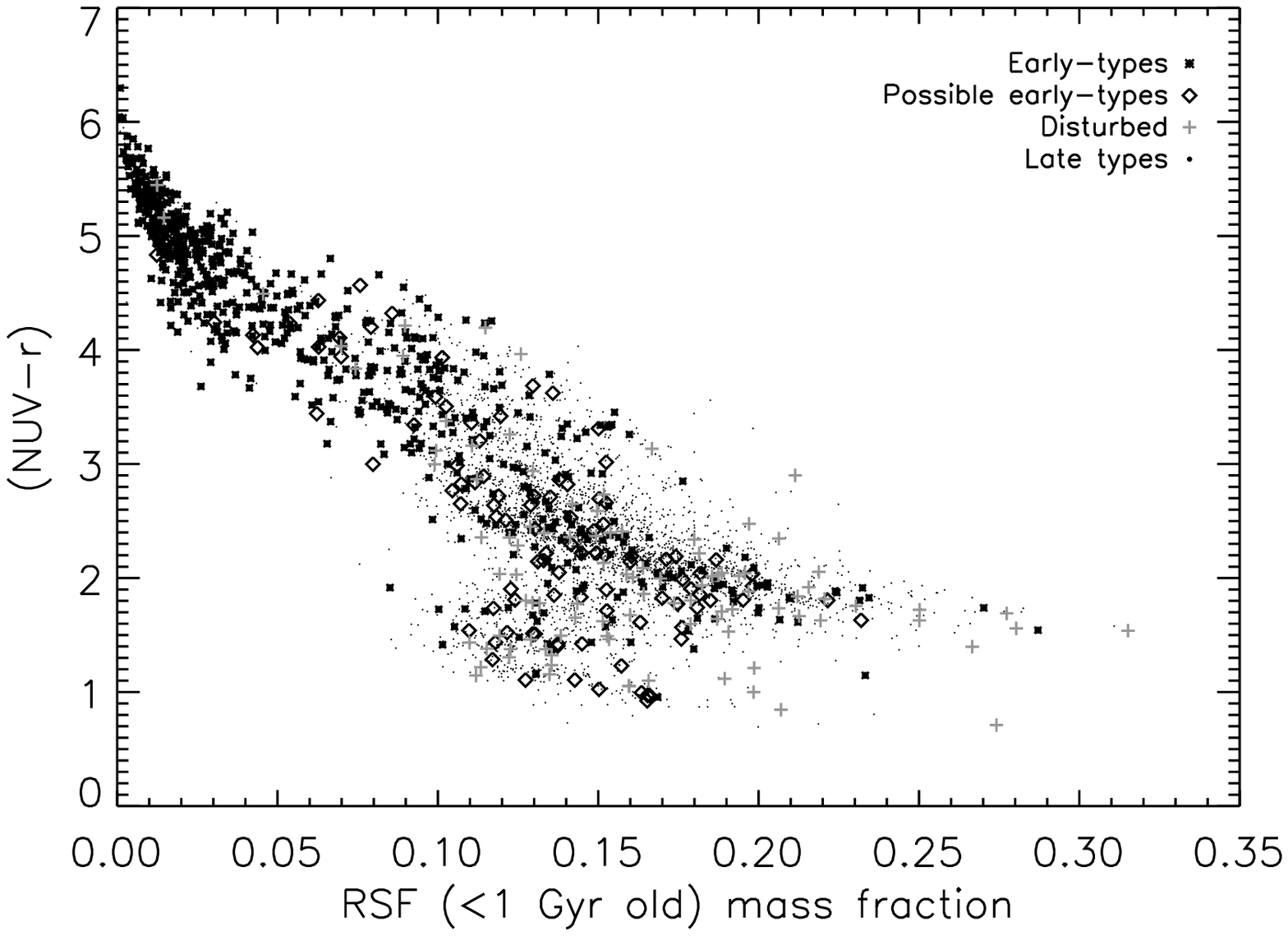}\\
\includegraphics[width=0.5\textwidth]{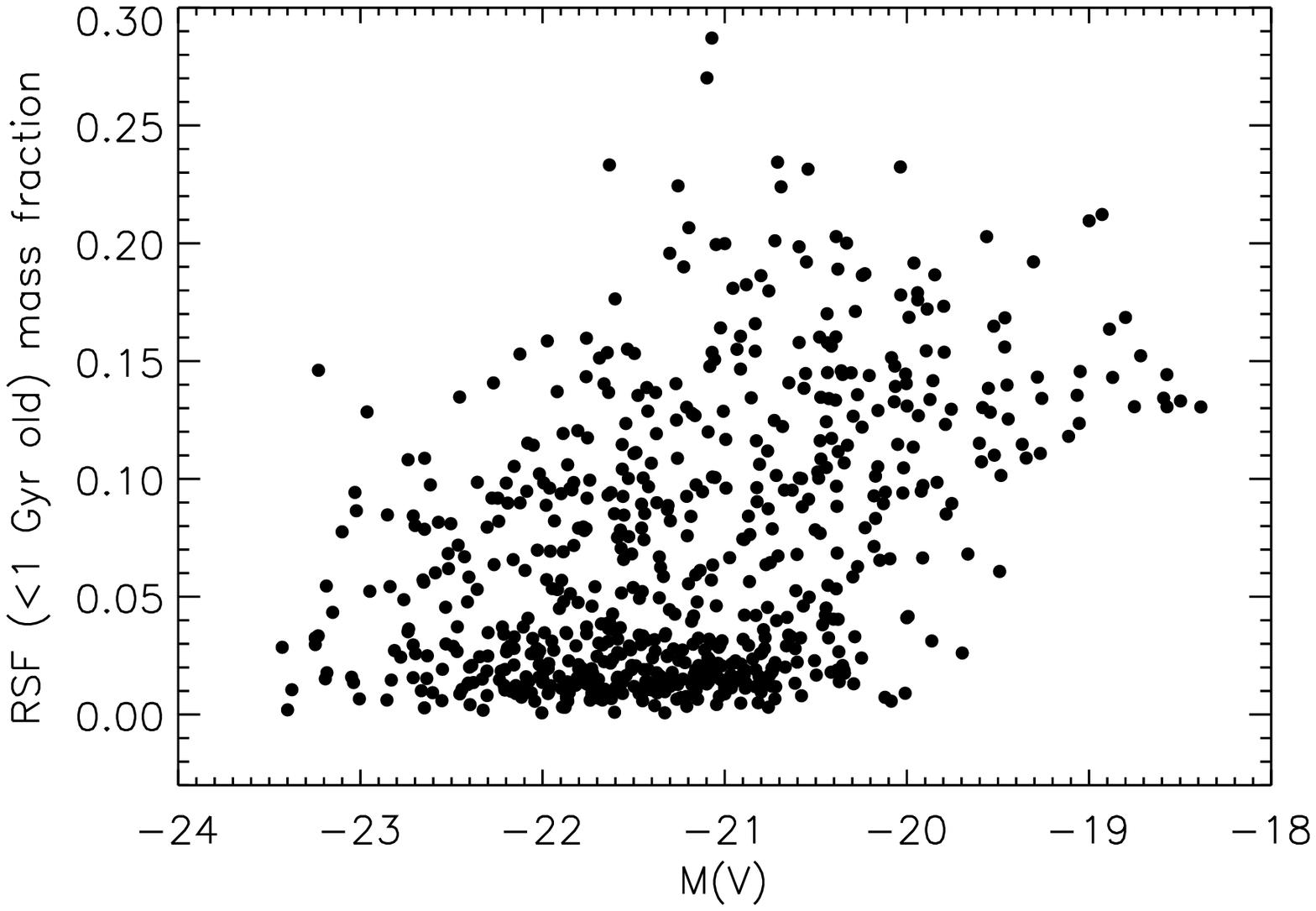} &
\includegraphics[width=0.5\textwidth]{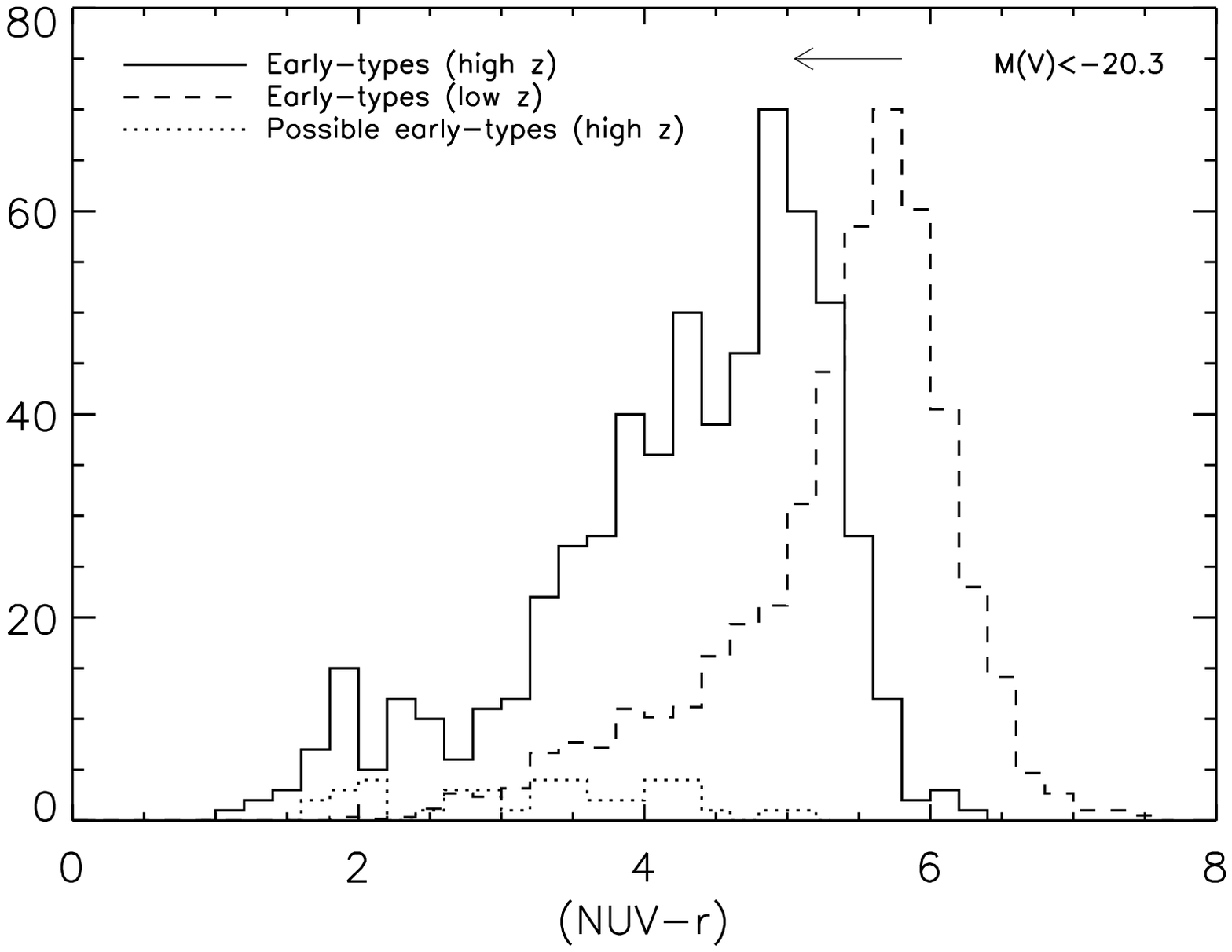}
\end{array}$
\caption{TOP LEFT: Rest-frame $(NUV-r)$ colour-magnitude relation
(top left) of the E-CDFS galaxy population. TOP RIGHT: Rest-frame
$(NUV-r)$ colour, plotted against the RSF in the E-CDFS galaxy
population. Note that here, RSF is explicitly defined as the mass
fraction in stars less than a Gyr old. BOTTOM LEFT: The RSF
plotted against absolute V-band magnitude for early-type galaxies
only. BOTTOM RIGHT: Comparison of the UV colours of the high
redshift early-type sample to their counterparts at low redshift
from the GALEX-SDSS work.} \label{fig:nuv_cmr}
\end{figure}

Using their GEMS images, K07 extracted a sample of early-type
galaxies by morphologically classifying a parent sample of
$\sim4500$ objects. Galaxy SFHs were estimated, and RSF values
extracted, by comparing the multi-wavelength photometry of each
observed galaxy with synthetic galaxy populations, generated in
the framework of the $\Lambda$CDM paradigm, using the
semi-analytical model of Khochfar \& Burkert\citep{Khochfar2003}.
Figure \ref{fig:nuv_cmr} shows the main results from this
analysis. It is apparent from the top left panel that the large
scatter observed in the NUV CMR in the local Universe (Figure
\ref{fig:opt_nuv_cmr}) persists in the intermediate redshift
Universe. Since the solid lines in this plot indicate the
positions of simple stellar populations of solar metallicity that
form at $z=2$, the colours of the early-type galaxy population
(shown in grey) suggest that a very small fraction of these
objects are consistent with \emph{purely} passive ageing since
high redshift. K07 estimated that $\sim1.1$ percent of early-types
in their sample are consistent (within errors) with \emph{purely}
passive ageing since $z=2$. This value drops to $\sim0.24$ percent
and $\sim0.15$ percent for $z=3$ and $z=5$ respectively. The
results of this study indicates that the luminous ($M(V)<-21$)
early-type population shows a typical RSF between 5 and 13\% in
the redshift range $0.5<z<1$, while early-types on the broad red
sequence ($NUV-r>4$) typically show RSF values less than 5\%
(bottom left panel in Figure \ref{fig:nuv_cmr}). The reddest
early-types (which are also the most luminous) are virtually
quiescent with RSF values of $\sim1$\%.

In contrast to their low-redshift counterparts, the early-type
population in E-CDFS shows a pronounced bimodality in the $NUV-r$
colour distribution, around $NUV-r\sim3$ (bottom right panel in
Figure \ref{fig:nuv_cmr}). The peak of the high-redshift $NUV-r$
distribution shows a relative shift from its low-redshift
counterpart which is close to what might be expected from the
passive ageing of a simple stellar population forming at high
redshift (shown by the arrow). This indicates that the bulk of the
stellar population in early-type galaxies, at least on the red
sequence, is overwhelmingly old. The blue peak in the
high-redshift $NUV-r$ distribution contains $\sim15$\% of the
early-type population, with an average RSF of $\sim11$\%. In
comparison, the bluest 15\% of the low-redshift early-type
population has an average RSF of $\sim6$\%, indicating that star
formation activity in the most active early- types has halved
between $z\sim0.7$ and present day. Finally, within the errors,
K07 found a weak trend of increasing RSF with redshift, from
$\sim7$\% at $z=0.5$ to $\sim13$\% at $z=1$, with a typical
uncertainty in the RSF of $\sim2.5$\%.

Since the timescale of this study is $\sim2.5$ Gyrs, a simple
extrapolation (from RSF values in Figure \ref{fig:nuv_cmr})
indicates that the \emph{bulk} of the luminous ($-23<M(V)<-21$)
early-type population may typically form up to 10-15 percent of
their mass after $z=1$ (with a tail to higher values), while their
less luminous ($M(V)>-21$) counterparts form 30-60 percent of
their mass in the same redshift range. These values are probably
overestimated since the intensity of star formation is seen to
decrease between $z=0$ and redshifts probed by this study. This
tail-end of star formation should exist in \emph{intermediate-age}
(3-8 Gyr old) stellar populations in early-type galaxies at
present day.

Finally, it is worth noting that the star formation experienced by
luminous early-types at late epochs is indirectly consistent with
the observed value of alpha-enhancements in these objects in the
local Universe. For example, if one assumes that the bulk of the
stellar mass indeed forms rapidly at high redshift, producing an
alpha-enhancement typical of a `monolithic-type' burst and that
the subsequent star formation has an alpha-enhancement that
reflects the solar value (due to its longer timescale), then the
quantity of RSF ($<15$\%) in these galaxies is insufficient to
perturb the alpha-enhancement from the monolithic value. The RSF
values derived from the UV studies are therefore consistent with
the observed alpha-enhancements of the luminous early-type
population in the local Universe.

\section{Sources of recent star formation in early-type galaxies}
Several plausible sources of low-level star formation exist, which
could, either individually or collectively, explain the broadness
of the UV CMR and the RSF values calculated in these systems.
Condensation from the extensive hot gas reservoirs hosted by
massive early-type galaxies may provide a plausible source of
(cold) gas that could fuel RSF in these systems. While feedback
sources (e.g. AGN and supernovae) might be expected to maintain
the temperature of the hot gas reservoir and evaporate infalling
cold material in the most massive haloes \citep{Binney2004}, this
process may not be fully efficient, allowing some gas condensation
to take place, which could then result in low-level star formation
as the cold gas settles in the potential well.

\begin{figure}
\begin{center}
$\begin{array}{cc}
\includegraphics[width=0.5\textwidth]{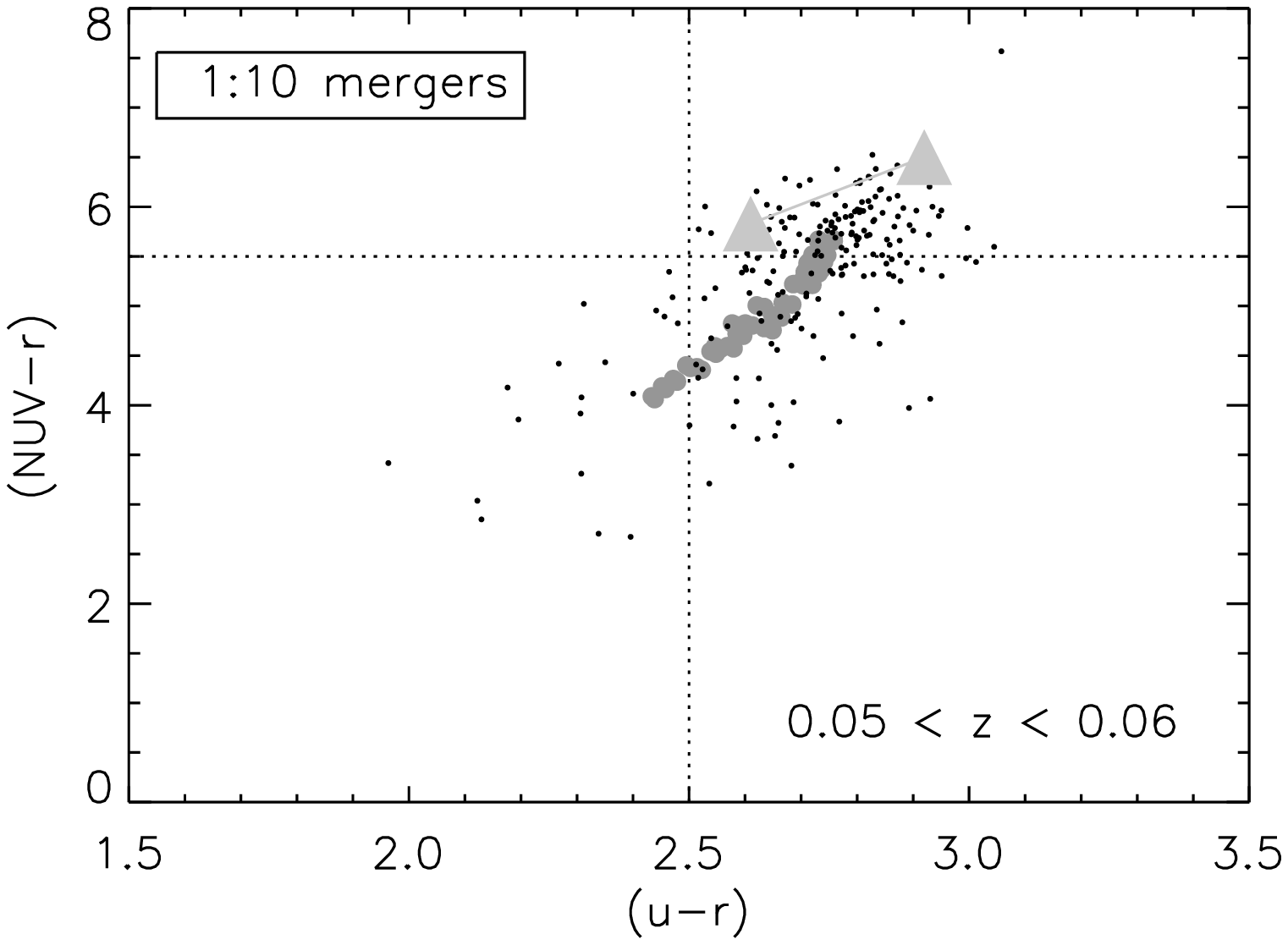}
\includegraphics[width=0.5\textwidth]{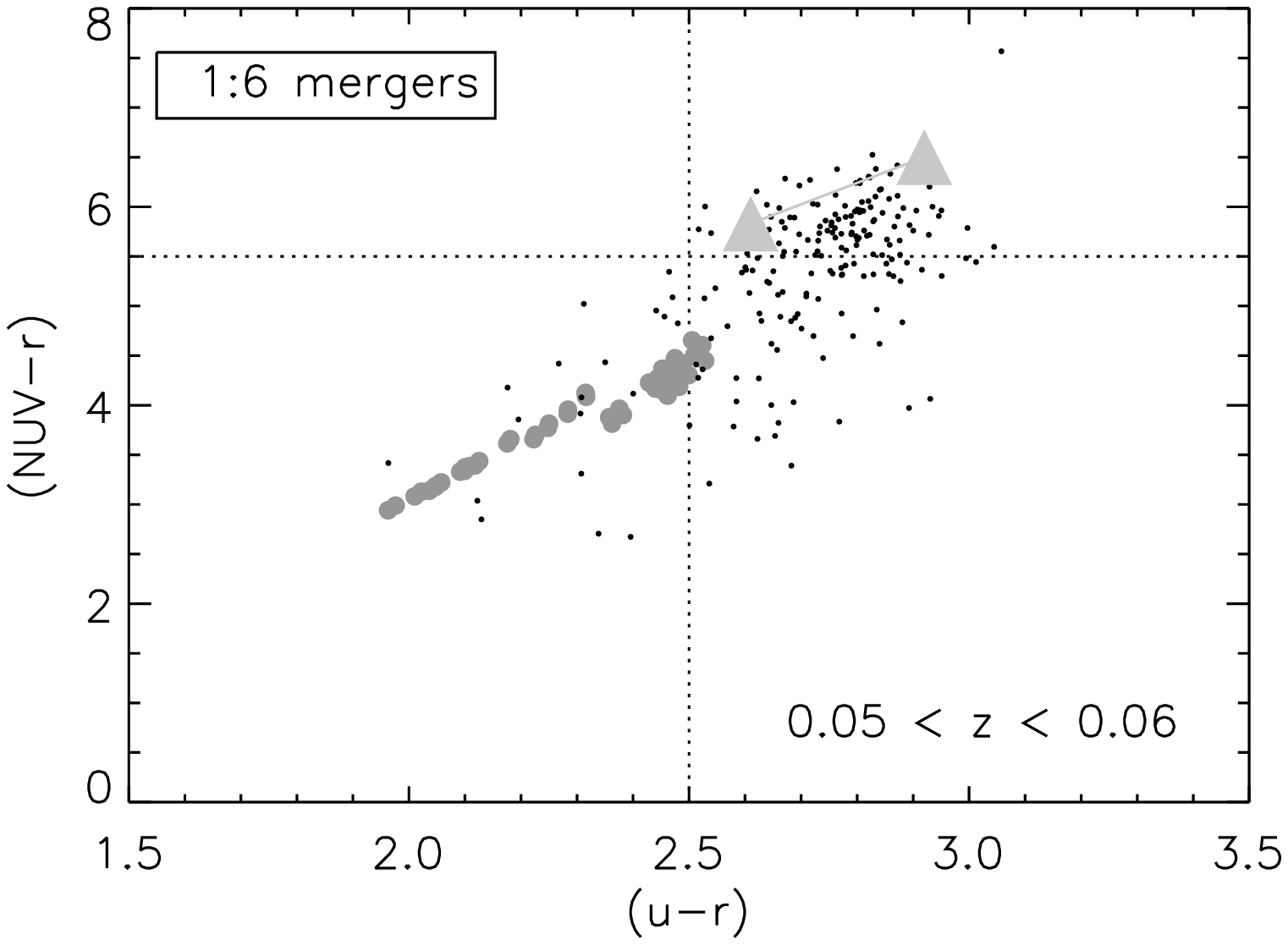}
\end{array}$
\caption{UV colours achieved by remnants of minor mergers (large
grey filled circles), where the mass ratios are 1:10 (left panel)
and 1:6 (right panel). The colours of the observed early-type
population is show using the small dots. Large triangles indicate
the position of a 9 Gyr SSP with half-solar and solar metallicity.
See the text in Section 4 for details.}
\label{fig:minor_mergers}
\end{center}
\end{figure}

Mergers and accretion events provide an alternative source of
young stars. The small levels of RSF in red sequence early-types
indicate that the star formation in these galaxies could be driven
either by accretion of small gas-rich satellites or through
\emph{largely} dry equal mass mergers. Simulations of binary
mergers at low redshift (Kaviraj et al., in prep) indicate that
the scatter in the early-type UV CMR can be reproduced by `minor'
mergers that have mass ratios between 1:6 and 1:10 and where the
accreted satellites have high gas fractions ($\gtrsim20$ percent).
Figure \ref{fig:minor_mergers} indicates the colours achieved by
the remnants of such minor mergers, where the mass ratios are 1:10
(left panel) and 1:6 (right panel). Note that the larger
progenitor is assumed to be a spheroidal galaxy (a pure stellar
bulge) while the satellite is modelled as a late-type object (a
pure disk). The tracks shown are from an ensemble of simulations,
where the free parameters are (a) the efficiency of star formation
(varied between 1\% and 10\%) (b) the gas fraction of the
satellite (varied between 20\% and 40\%) (c) the metallicity of
the satellite (varied between 0.004 and 0.04) and (d) the age of
the satellite (varied between 1 and 9 Gyrs). The mass, metallicity
and age of the parent spheroid are assumed to be
$10^{11}M_{\odot}$, 0.02 dex and 9 Gyrs respectively. The remnant
is `observed' at the point where the satellite finally disappears
into the parent spheroid so that images (e.g. through an SDSS
$r$-band filter) would indicate a single spheroidal object. The
colours shown are therefore the bluest possible for each scenario,
since star formation declines after this `final plunge' and the
remnant would be expected to redden by about 0.5-1 mag every Gyr.

Figure \ref{fig:minor_mergers} indicates that minor mergers, with
reasonable assumptions for the star formation prescriptions and
properties of the merger progenitors, can reproduce the entire
extent of the NUV CMR observed at low redshift. However, the
\emph{likelihood} of the minor merger channel clearly depends on
the expected frequency of such events. Merger statistics in the
$\Lambda$CDM model indicate that $\sim5-7$\% of luminous
early-types at $z\sim0$ would experience minor mergers with mass
ratios between 1:6 and 1:10 (Sadegh Khochfar, private
communication) within the last 0.5 Gyrs. Although further
investigation of this issue is necessary, it appears likely that
the broad red sequence ($NUV-r>5.8$) is composed of galaxies which
have not had any interaction within the last Gyr and that the
$\Lambda$CDM merger rates could plausibly account for most but not
all of the blue ($NUV-r<5.5$) galaxies in the early-type
population. This, in turn, implies that accretion from the halo
might play a role, as might other factors such as low dust
contents in star forming regions.

\section{Summary}
The advent of the GALEX space telescope and deep optical surveys
have revolutionised our understanding of the properties of the
early-type galaxy population in the rest-frame UV. The sensitivity
of the UV to recent star formation provides an unprecedented
opportunity to quantify the SFHs of early-type galaxies over the
last 8 billion years and put strong constraints on models for
their formation.

The first generation of studies that have harnessed the UV to
directly constrain the RSF in the early-type population indicate
that the early-type population, in general, forms stars over the
lifetime of the Universe. The SFH of luminous ($M(V)<-21$)
early-types is \emph{quasi-monolithic}, with more than 80\% of
their stellar mass forming at $z>1$.  Fainter early-types
experience more prolonged star formation, potentially forming a
substantial fraction (30-60\%) of their stellar mass at recent
epochs ($z<1$).The derived values of RSF are consistent with the
observed values of alpha-enhancements of luminous early-type
galaxies in the local Universe.

The rest-frame UV CMR shows a typical scatter of several
magnitudes over the entire redshift range $0<z<1$, indicating that
low-level recent star formation is widespread in the early-type
population. The intensity of recent star formation and the
\emph{bulk} of the UV colour distribution is consistent with what
might be expected from simulations of minor mergers (mass ratios
$\lesssim$ 1:6) in a $\Lambda$CDM cosmology.

Forthcoming studies with COSMOS\citep{COSMOS} and DEEP2
\citep{Davis2003} data will significantly consolidate our
understanding of the UV properties of the high-redshift early-type
population, alleviate the effect of cosmic variance and allow us
to the split the observed sample of galaxies not only by
luminosity (mass) but also by environment (using spectroscopic
redshifts), which is a key driver of galaxy
evolution\citep{Kaviraj2006a}.


\section{Acknowledgements}
\small This research was supported by a Leverhulme Early Career
Fellowship, a BIPAC fellowship at the University of Oxford and a
Research Fellowship at Worcester College, Oxford. Joe Silk,
Ignacio Ferreras and Roger Davies are thanked for a careful
reading of the manuscript and many useful comments. The work
presented in this review was performed in collaboration with
Sukyoung Yi, Suk-Jin Yoon, Julien Devriendt, Ignacio Ferreras,
Kevin Schawinski, Sadegh Khochfar, Joe Silk, Eric Gawiser, Pieter
van Dokkum, the GALEX Science Team and the MUSYC collaboration. I
thank Sebastien Peirani for allowing me to use the results of the
minor merger simulations. Rachel Somerville, Nick Scoville, Chris
Wolf, Daniel Thomas and Claudia Maraston are thanked for useful
discussions.
\normalsize


\bibliographystyle{unsrt}
\bibliography{references}

\begin{thebibliography}{10}

\bibitem{ELS62}
O.~J. {Eggen}, D.~{Lynden-Bell}, and A.~R. {Sandage}.
\newblock {\em ApJ}, 136:748, 1962.

\bibitem{Larson74}
R.~B. {Larson}.
\newblock {\em MNRAS}, 166:385, 1974.

\bibitem{Chiosi2002}
C.~{Chiosi} and G.~{Carraro}.
\newblock {\em MNRAS}, 335:335, 2002.

\bibitem{Jorg1996}
I.~{Jorgensen}, M.~{Franx}, and P.~{Kjaergaard}.
\newblock {\em MNRAS}, 280:167--185, 1996.

\bibitem{Saglia1997}
R.~P. {Saglia}, M.~{Colless}, G.~{Baggley}, E.~{Bertschinger}, D.~{Burstein},
  R.~L. {Davies}, R.~K. {McMahan}, and G.~{Wegner}.
\newblock {The EFAR Fundamental Plane}.
\newblock In M.~{Arnaboldi}, G.~S. {Da Costa}, and P.~{Saha}, editors, {\em ASP
  Conf. Ser. 116: The Nature of Elliptical Galaxies; 2nd Stromlo Symposium},
  pages 180--+, 1997.

\bibitem{Forbes1998}
D.~A. {Forbes}, T.~J. {Ponman}, and R.~J.~N. {Brown}.
\newblock {\em ApJ}, 508:L43--L46, 1998.

\bibitem{Peebles2002}
P.~J.~E. {Peebles}.
\newblock In {\em ASP Conf. Ser. 283: A New Era in Cosmology}, pages 351--+,
  2002.

\bibitem{Franx1993}
M.~{Franx}.
\newblock {\em PASP}, 105:1058--1062, 1993.

\bibitem{Franx1995}
M.~{Franx}.
\newblock {Measuring the Evolution of the M/L Ratio from the Fundamental
  Plane}.
\newblock In P.~C. {van der Kruit} and G.~{Gilmore}, editors, {\em IAU Symp.
  164: Stellar Populations}, pages 269--+, 1995.

\bibitem{VD1996}
P.~G. {van Dokkum} and M.~{Franx}.
\newblock {\em MNRAS}, 281:985--1000, 1996.

\bibitem{BLE92}
R.~G. {Bower}, J.~R. {Lucey}, and R.~{Ellis}.
\newblock {\em MNRAS}, 254:589, 1992.

\bibitem{Bender1997}
R.~{Bender}.
\newblock {Structure; Formation and Ages of Elliptical Galaxies}.
\newblock In {\em ASP Conf. Ser. 116: The Nature of Elliptical Galaxies; 2nd
  Stromlo Symposium}, pages 11--+, 1997.

\bibitem{Ellis97}
R.~S. {Ellis}, I.~{Smail}, A.~{Dressler}, W.~J. {Couche}, A.~Jr. {Oemler},
  H.~{Butcher}, and R.~M. {Sharples}.
\newblock {\em ApJ}, 483:582, 1997.

\bibitem{Stanford98}
S.~A. {Stanford}, P.~R.~M. {Eisenhardt}, and M.~{Dickinson}.
\newblock {\em ApJ}, 492:461, 1998.

\bibitem{Gladders98}
M.~D. {Gladders}, O.~{Lopez-Cruz}, H.~K.~C. {Yee}, and T.~{Kodama}.
\newblock {\em ApJ}, 501:571, 1998.

\bibitem{VD2000}
P.~G. {van Dokkum}, M.~{Franx}, D.~{Fabricant}, G.~D. {Illingworth}, and D.~D.
  {Kelson}.
\newblock {\em ApJ}, 541:95--111, 2000.

\bibitem{Thomas1999}
D.~{Thomas}, L.~{Greggio}, and R.~{Bender}.
\newblock {\em MNRAS}, 302:537--548, 1999.

\bibitem{Greggio1983}
L.~{Greggio} and A.~{Renzini}.
\newblock {\em A\&A}, 118:217--222, 1983.

\bibitem{Toomre_mergers}
A.~{Toomre}.
\newblock In B.~M. {Tinsley} and R.~B. {Larson}, editors, {\em Evolution of
  Galaxies and Stellar Populations}, pages 401--+, 1977.

\bibitem{Barnes1992a}
J.~E. {Barnes} and L.~{Hernquist}.
\newblock {\em ARAA}, 30:705--742, 1992.

\bibitem{Barnes1992b}
J.~E. {Barnes} and L.~{Hernquist}.
\newblock {\em Nature}, 360:715--717, 1992.

\bibitem{Hernquist1993}
L.~{Hernquist}.
\newblock In J.~M. {Shull} and H.~A. {Thronson}, editors, {\em ASSL Vol. 188:
  The Environment and Evolution of Galaxies}, pages 327--+, 1993.

\bibitem{Barnes1996}
J.~E. {Barnes} and L.~{Hernquist}.
\newblock {\em ApJ}, 471:115--+, 1996.

\bibitem{Bender1996}
R.~{Bender}.
\newblock In R.~{Bender} and R.~L. {Davies}, editors, {\em IAU Symp. 171: New
  Light on Galaxy Evolution}, pages 181--+, 1996.

\bibitem{Cole2000}
S.~{Cole}, C.~G. {Lacey}, C.~M. {Baugh}, and C.~S. {Frenk}.
\newblock {\em MNRAS}, 319:168--204, 2000.

\bibitem{Hatton2003}
S.~{Hatton}, J.~E.~G. {Devriendt}, S.~{Ninin}, F.~R. {Bouchet},
  B.~{Guiderdoni}, and D.~{Vibert}.
\newblock {\em MNRAS}, 343:75--106, 2003.

\bibitem{Khochfar2003}
S.~{Khochfar} and A.~{Burkert}.
\newblock {\em ApJ}, 597:L117--L120, 2003.

\bibitem{K96}
G.~{Kauffmann}.
\newblock {\em MNRAS}, 281:487--492, 1996.

\bibitem{K98}
G.~{Kauffmann} and S.~{Charlot}.
\newblock {\em MNRAS}, 294:705--+, 1998.

\bibitem{Sadler1985}
E.~M. {Sadler} and O.~E. {Gerhard}.
\newblock 214:177--187, 1985.

\bibitem{Tomita2000}
A.~{Tomita}, K.~{Aoki}, M.~{Watanabe}, T.~{Takata}, and S.-i. {Ichikawa}.
\newblock {\em AJ}, 120:123--130, 2000.

\bibitem{Tran2001}
H.~D. {Tran}, Z.~{Tsvetanov}, H.~C. {Ford}, J.~{Davies}, W.~{Jaffe}, F.~C. {van
  den Bosch}, and A.~{Rest}.
\newblock {\em AJ}, 121:2928--2942, 2001.

\bibitem{sauron5}
M.~{Sarzi}, J.~{Falc{\'o}n-Barroso}, R.~L. {Davies}, R.~{Bacon}, M.~{Bureau},
  M.~{Cappellari}, P.~T. {de Zeeuw}, E.~{Emsellem}, K.~{Fathi},
  D.~{Krajnovi{\'c}}, H.~{Kuntschner}, R.~M. {McDermid}, and R.~F. {Peletier}.
\newblock {\em MNRAS}, 366:1151--1200, March 2006.

\bibitem{Malin83}
D.~F. {Malin} and D.~{Carter}.
\newblock {\em ApJ}, 274:534--540, 1983.

\bibitem{VD2005}
P.~G. {van Dokkum}.
\newblock {\em AJ}, 130:2647--2665, 2005.

\bibitem{sauron2}
P.~T. {de Zeeuw}, M.~{Bureau}, E.~{Emsellem}, R.~{Bacon}, C.~M. {Carollo},
  Y.~{Copin}, R.~L. {Davies}, H.~{Kuntschner}, B.~W. {Miller}, G.~{Monnet},
  R.~F. {Peletier}, and E.~K. {Verolme}.
\newblock {\em MNRAS}, 329:513--530, 2002.

\bibitem{Trager2000a}
S.~C. {Trager}, S.~M. {Faber}, G.~{Worthey}, and J.~J. {Gonz{\' a}lez}.
\newblock {\em AJ}, 119:1645--1676, 2000.

\bibitem{Trager2000b}
S.~C. {Trager}, S.~M. {Faber}, G.~{Worthey}, and J.~J. {Gonz{\' a}lez}.
\newblock {\em AJ}, 120:165--188, 2000.

\bibitem{Fukugita2004}
M.~{Fukugita}, O.~{Nakamura}, E.~L. {Turner}, J.~{Helmboldt}, and R.~C.
  {Nichol}.
\newblock {\em ApJL}, 601:L127--L130, 2004.

\bibitem{Menanteau2001a}
F.~{Menanteau}, R.~G. {Abraham}, and R.~S. {Ellis}.
\newblock {\em MNRAS}, 322:1--12, 2001.

\bibitem{Ferreras2005}
I.~{Ferreras}, T.~{Lisker}, C.~M. {Carollo}, S.~J. {Lilly}, and B.~{Mobasher}.
\newblock {\em ApJ}, 635:243--259, 2005.

\bibitem{Ellis2001}
R.~S. {Ellis}, R.~G. {Abraham}, and M.~{Dickinson}.
\newblock {\em ApJ}, 551:111--130, 2001.

\bibitem{VD2001a}
P.~G. {van Dokkum} and M.~{Franx}.
\newblock {\em ApJ}, 553:90--102, 2001.

\bibitem{Kaviraj2005a}
S.~{Kaviraj}, J.~E.~G. {Devriendt}, I.~{Ferreras}, and S.~K. {Yi}.
\newblock {\em MNRAS}, 360:60--68, 2005.

\bibitem{Dressler80}
A.~{Dressler}.
\newblock {\em ApJ}, 236:351--365, 1980.

\bibitem{BO84}
H.~{Butcher} and A.~{Oemler}.
\newblock {\em ApJ}, 285:426--438, 1984.

\bibitem{Dressler97}
A.~{Dressler}, A.~J. {Oemler}, W.~J. {Couch}, I.~{Smail}, R.~S. {Ellis},
  A.~{Barger}, H.~{Butcher}, B.~M. {Poggianti}, and R.~M. {Sharples}.
\newblock {\em ApJ}, 490:577, 1997.

\bibitem{Couch98}
W.~J. {Couch}, A.~J. {Barger}, I.~{Smail}, R.~S. {Ellis}, and R.~M. {Sharples}.
\newblock {\em ApJ}, 497:188--+, 1998.

\bibitem{Margoniner2001}
V.~E. {Margoniner}, R.~R. {de Carvalho}, R.~R. {Gal}, and S.~G. {Djorgovski}.
\newblock {\em ApJL}, 548:L143--L146, 2001.

\bibitem{Andreon2004}
S.~{Andreon}, C.~{Lobo}, and A.~{Iovino}.
\newblock {\em MNRAS}, 349:889--898, April 2004.

\bibitem{Borch2006}
A.~{Borch}, K.~{Meisenheimer}, E.~F. {Bell}, H.-W. {Rix}, C.~{Wolf}, S.~{Dye},
  M.~{Kleinheinrich}, Z.~{Kovacs}, and L.~{Wisotzki}.
\newblock {\em A\&A}, 453:869--881, 2006.

\bibitem{Bundy2006}
K.~{Bundy}, R.~S. {Ellis}, C.~J. {Conselice}, J.~E. {Taylor}, M.~C. {Cooper},
  C.~N.~A. {Willmer}, B.~J. {Weiner}, A.~L. {Coil}, K.~G. {Noeske}, and
  P.~R.~M. {Eisenhardt}.
\newblock {\em ApJ}, 651:120--141, 2006.

\bibitem{VD99}
P.~G. {van Dokkum}, M.~{Franx}, D.~{Fabricant}, D.~D. {Kelson}, and
  {Illingworth}.
\newblock {\em ApJ}, 520:L95, 1999.

\bibitem{Bell2004}
E.~F. {Bell}, C.~{Wolf}, K.~{Meisenheimer}, H.-W. {Rix}, A.~{Borch}, S.~{Dye},
  M.~{Kleinheinrich}, L.~{Wisotzki}, and D.~H. {McIntosh}.
\newblock {\em ApJ}, 608:752--767, 2004.

\bibitem{Faber2007}
S.~M. {Faber} and {DEEP2 collaboration}.
\newblock {\em ApJ}, 665:265--294, 2007.

\bibitem{SDSSDR4}
J.~K. {Adelman-McCarthy} and {SDSS collaboration}.
\newblock {\em ApJS}, 162:38--48, 2006.

\bibitem{Davis2003}
M.~{Davis} and {DEEP2 collaboration}.
\newblock In {\em Discoveries and Research Prospects from 6- to 10-Meter-Class
  Telescopes II. Edited by Guhathakurta, Puragra. Proceedings of the SPIE,
  Volume 4834, pp. 161-172 (2003).}, volume 4834, pages 161--172, 2003.

\bibitem{Worthey1994}
G.~{Worthey}.
\newblock {\em ApJS}, 95:107--149, 1994.

\bibitem{Kaviraj2006c}
S.~{Kaviraj}, S.-C. {Rey}, R.~M. {Rich}, Y.~{Lee}, S.-J. {Yoon}, and S.~K.
  {Yi}.
\newblock {\em MNRAS in press; astro-ph/0601050}, 2006.

\bibitem{Martin2005}
D.~C. {Martin} and {GALEX collaboration}.
\newblock {\em ApJ}, 619:L1--L6, 2005.

\bibitem{Ferreras2000}
I.~{Ferreras} and J.~{Silk}.
\newblock {\em ApJL}, 541:L37--L40, 2000.

\bibitem{Ferreras2002}
I.~{Ferreras}, E.~{Scannapieco}, and J.~{Silk}.
\newblock {\em ApJ}, 579:247--260, 2002.

\bibitem{Yi97}
S.~{Yi}, P.~{Demarque}, and A.~J. {Oemler}.
\newblock {\em ApJ}, 486:201--+, 1997.

\bibitem{Yi99}
S.~{Yi}, Y.-W. {Lee}, J.-H. {Woo}, J.-H. {Park}, P.~{Demarque}, and A.~J.
  {Oemler}.
\newblock {\em ApJ}, 513:128--141, 1999.

\bibitem{Yi2005}
S.~K. {Yi}, S.-J. {Yoon}, S.~{Kaviraj}, J.-M. {Deharveng}, and {the GALEX
  Science Team}.
\newblock {\em ApJ}, 619:L111--L114, 2005.

\bibitem{Rich2005}
R.~M. {Rich}, S.~{Salim}, J.~{Brinchmann}, S.~{Charlot}, and {the GALEX
  collaboration}.
\newblock {\em ApJ}, 619:L107--L110, 2005.

\bibitem{Bernardi2003d}
M.~{Bernardi} and {the SDSS collaboration}.
\newblock {\em AJ}, 125:1882--1896, 2003.

\bibitem{Kaviraj2006b}
S.~{Kaviraj} and {GALEX Science Team}.
\newblock {\em ApJ in press - to appear in GALEX dedicated issue in Dec 2007
  (astro-ph/0601036)}.

\bibitem{Baldwin1981}
J.~A. {Baldwin}, M.~M. {Phillips}, and R.~{Terlevich}.
\newblock {\em PASP}, 93:5--19, 1981.

\bibitem{Kauffmann2003}
G.~{Kauffmann} and {SDSS collaboration}.
\newblock {\em MNRAS}, 346:1055--1077, 2003.

\bibitem{Kaviraj2007a}
S.~{Kaviraj}, S.~{Khochfar}, K.~{Schawinski}, S.~K. {Yi}, E.~{Gawiser},
  J.~{Silk}, S.~N. {Virani}, C.~{Cardamone}, P.~G. {van Dokkum}, and C.~M.
  {Urry}.
\newblock {\em astro-ph/0709.0806}, 709, 2007.

\bibitem{Gawiser2006}
E.~{Gawiser} and {MUSYC collaboration}.
\newblock {\em ApJS}, 162:1--19, 2006.

\bibitem{Wolf2004}
C.~{Wolf}, K.~{Meisenheimer}, M.~{Kleinheinrich}, A.~{Borch}, S.~{Dye},
  M.~{Gray}, L.~{Wisotzki}, E.~F. {Bell}, H.-W. {Rix}, A.~{Cimatti},
  G.~{Hasinger}, and G.~{Szokoly}.
\newblock {\em A\&A}, 421:913--936, 2004.

\bibitem{Rix2004}
H.-W. {Rix} and {GEMS collaboration}.
\newblock {\em ApJS}, 152:163--173, 2004.

\bibitem{Binney2004}
J.~{Binney}.
\newblock {\em MNRAS}, 347:1093--1096, 2004.

\bibitem{COSMOS}
N.~{Scoville} and {COSMOS collaboration}.
\newblock {\em ApJS}, 172:38--45, 2007.

\bibitem{Kaviraj2006a}
S.~{Kaviraj}, J.~E.~G. {Devriendt}, I.~{Ferreras}, S.~K. {Yi}, and J.~{Silk}.
\newblock {\em MNRAS; astro-ph/0602347}, 2006.

\end{thebibliography}


\end{document}